\documentclass[12pt,a4paper]{article}
 \usepackage{natbib}
\usepackage[colorlinks,citecolor=blue,urlcolor=blue,filecolor=blue,backref=page]{hyperref}
\usepackage{mathptmx}       % selects Times Roman as basic font
\usepackage{helvet}         % selects Helvetica as sans-serif font
\usepackage{courier}        % selects Courier as typewriter font
\usepackage{makeidx}         % allows index generation
\usepackage{graphicx}        % standard LaTeX graphics tool
\usepackage{subcaption} % the subfigure and subfig packages are deprecated
\usepackage{multicol}        % used for the two-column index
\usepackage[bottom]{footmisc}% places footnotes at page bottom
\usepackage{amsmath,amssymb,amsfonts,bm}        % AMS math packages
\makeindex             % used for the subject index
\title{Bayesian Modelling and Quantification of\\Raman Spectroscopy}
\author{Matthew T. Moores\footnote{School of Engineering Sciences, Lappeenranta-Lahti University of Technology, Yliopistonkatu 34, 53850 Lappeenranta, Finland.} 
\and  Kirsten Gracie\footnote{Department of Pure \& Applied Chemistry, University of Strathclyde, Glasgow G1 1RD, UK.} \and Jake Carson\footnote{Mathematics Institute, University of Warwick, Coventry CV4 7AL, UK.} \and Karen Faulds\footnotemark[2] \and Duncan Graham\footnotemark[2] \and Mark Girolami \footnote{Department of Engineering, University of Cambridge, Cambridge CB2 1PZ, UK.}\; \footnote{Alan Turing Institute, London NW1 2BD, UK.}}
\date{}

\begin{document}
%\footnotetext[1]
%\footnotetext[2]
%\footnotetext[3]
%\footnotetext[4]
%\footnotetext[5]
\maketitle

\abstract{Raman spectroscopy can be used to identify molecules such as DNA by
the characteristic scattering of light from a laser. It is sensitive at very low concentrations
and can accurately quantify the amount of a given molecule in a sample. The
presence of a large, nonuniform background presents a major challenge to analysis
of these spectra. To overcome this challenge, we introduce a sequential Monte Carlo (SMC) algorithm to separate
the observed spectrum into a series of peaks plus a smoothly-varying baseline,
corrupted by additive white noise. The peaks are modelled using  Lorentzian or Gaussian broadening functions, while the baseline is estimated using a penalised cubic
spline. This latent continuous representation accounts for differences in resolution
between measurements. By incorporating this representation in a Bayesian model, we can quantify the relationship between molecular concentration and peak
intensity, thereby providing an improved estimate of the limit of detection (LOD), which is of major importance in analytical chemistry.
}

\section{Introduction}
\label{sec:intro}
Understanding the chemical composition of mineral and biological samples is vital for many applications, including the search for evidence of past life on Mars \citep{Hutchinson2014,Maurice2021} and medical diagnostics of cancer and other diseases \citep{Laing2017,braddick2025emerging}. Spectroscopy is a measurement technique that can be used to estimate compositions by observing the interaction of matter and electromagnetic radiation, such as light from a laser. In the case of Raman spectroscopy, this produces a complex pattern of peaks that correspond to the vibrational modes of the molecules in the sample. The spectral signature produced by Raman scattering is highly specific, enabling simultaneous identification and quantification of several molecules in a multiplex \citep{Zhong2011,Gracie2014}.

% For figures use
%
\begin{figure}[ht]
\centering
\includegraphics[width=\textwidth]{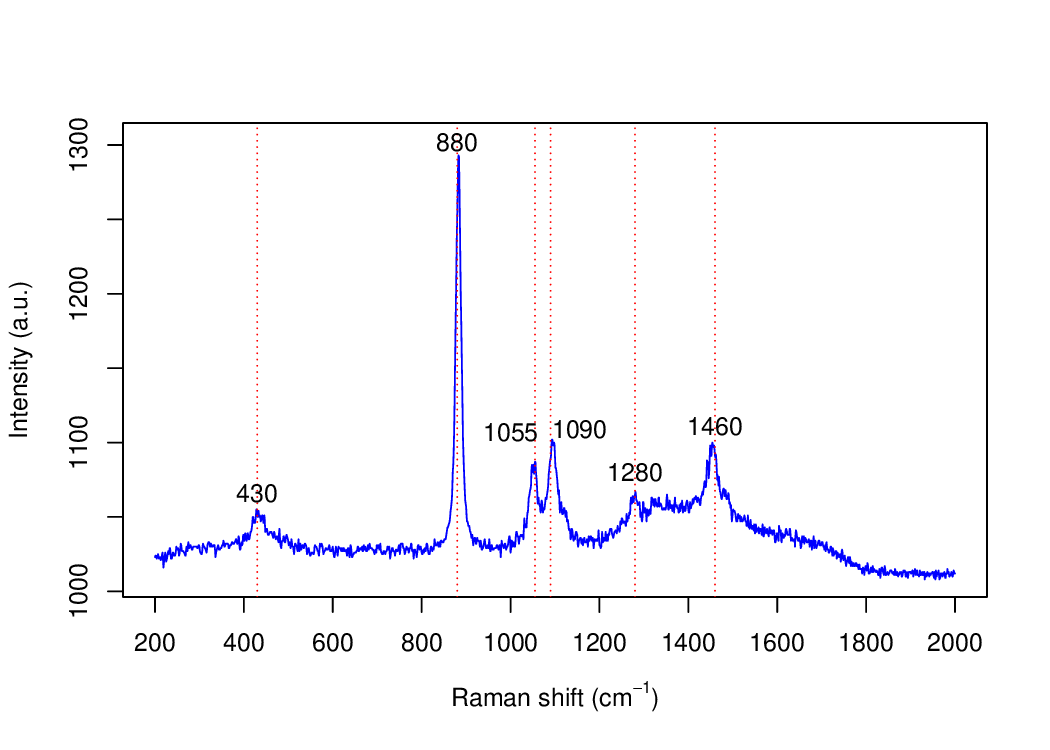}
\caption{Raman spectrum of ethanol (\textsf{EtOH}), showing the locations of 6 major peaks (430, 880, 1055, 1090, 1280 \& 1460 cm$^{-1}$).}
\label{f:EtOH}
\end{figure}

An example Raman spectrum for ethanol (\textsf{EtOH}) is shown in Fig.~\ref{f:EtOH}. Here we are focused on the fingerprint region for organic molecules: wavenumbers $\Delta \tilde{\nu}$ in the range 200 to 2000 cm$^{-1}$. The vertical axis is measured in  arbitrary units (a.u.), since the observed signal intensity is dependent on the calibration of the spectrometer, among several other factors. Ethanol is a relatively simple molecule, in comparison to the Raman-active dye labels that are analysed in Sect.~\ref{s:results}. The 6 major peaks in its spectral signature can be directly attributed to vibrational modes of the bonds between its 9 atoms \citep{Mammone1980,Lin-Vien1991}. Most of these peaks are well-separated, so that the shape of the smooth baseline function can be readily discerned. %The number and locations of these peaks are extremely stable, so much so that EtOH can be used as a calibration standard for  Raman offsets.

A further advantage of Raman spectroscopy is that the amplitudes of the peaks increase linearly with the concentration of the molecule \citep{Jones1999}. A dilution study can be performed to measure Raman spectra at a range of known concentrations. By fitting a linear regression model to this data, it is then possible to estimate the limit of detection (LOD) for the molecule, which is the minimum concentration that the Raman peaks can be distinguished from noise:
\begin{equation}\label{eq:lod}
c_{LOD} = \frac{3 \sigma_\epsilon}{\beta_p}
\end{equation}
where $\sigma_\epsilon$ is the standard deviation of the additive white noise and $\beta_p$ is the linear regression coefficient for peak $p$ that measures the relationship between amplitude and concentration. The LOD is usually only estimated for the single, largest peak (univariate calibration). For example, the LOD for the \textsf{EtOH} peak at 880 cm$^{-1}$ has been estimated as 1.2 millimolar (mM) concentration \citep{Boyaci2012}. This can be used to estimate the alcohol content of commodity spirits, such as whisky, vodka, or gin, as well as to detect counterfeits \citep{Ellis2017}.

More complex molecules might not have a single, dominant peak that is well-separated from the others. In this case, multivariate calibration (MVC) can be used to quantify several peaks simultaneously \citep{Pelletier2003,Varmuza2009}. Traditional chemometric methods for MVC include direct classical least squares \citep[DCLS;][]{Haaland1980} and partial least squares \citep[PLS;][]{Wold2001,Frank1993}. However, these methods rely on accurate baseline subtraction as a data pre-processing step. 

Existing approaches to automated baseline correction include asymmetric least squares \citep{Boelens2005,He2014}, iterative polynomial fit \citep{Gan2006,Lieber2003}, locally weighted smoothing \citep{Ruckstuhl2001}, and wavelet decomposition \citep{Cai2001,Galloway2009}. See \cite{Schulze2005,Liland2010} for comparative reviews. Subtracting the baseline as a pre-processing step ignores the uncertainty in the estimate, and can introduce artefacts that cause bias in the resulting quantification. The remaining signal can have very low likelihood, once the shape of the peaks is taken into account. This is particularly a problem for Raman spectra of complex molecules, where the peaks overlap to such a degree that the baseline is seldom directly observed. %Clearly it would be preferable to estimate the parameters of both the peaks and the baseline jointly.

%An iterative algorithm for estimation of the baseline and peaks was introduced by \cite{deRooi2012}. They combined a penalised spline for the baseline with a mixture model to differentiate between the peaks and the residual noise. The noise was assumed to be Gaussian, while the peaks were represented using a uniform distribution on the positive real numbers. This nonparametric model does not assume any functional form for the broadening of the peaks. Thus, it cannot be used to estimate quantities of scientific interest, such as the LOD in \eqref{eq:lod}. Treating the peaks as an unordered collection of points also ignores dependence between neighbouring values in the spectrum and makes it difficult to compare spectra with different resolutions.

% \citeauthor{deRooi2012} fit their model using an expectation-maximisation (EM) algorithm \citep{Dempster1977}, which is known to be highly sensitive to its initial conditions and to be prone to converging to suboptimal local maxima \citep[ch. 3]{McLachlan2008}. These problems can be mitigated using multiple, random starting points combined with annealed likelihood \citep{Zhou2010}, but it would be preferable to use an algorithm that explores the full parameter space.

Alternatively, each peak could be represented as a continuous function, in accordance with the known physical properties of Raman spectroscopy. Doppler broadening is a result of the emitted photons being red (blue) shifted due to particles moving away from (towards) the sensor. Since the particles are undergoing Brownian motion, this gives rise to a Gaussian function:
\begin{equation}\label{eq:GaussPk}
f_G(\nu_j \mid \ell_p, \varphi_p) \propto \exp\left\{ - \frac{(\nu_j - \ell_p)^2}{2 \varphi_p^2} \right\}
\end{equation}
where $\nu_j$ is the $j$th wavenumber in the spectrum, $\ell_p$ is the location of peak $p$, and $\varphi_p$ is a scale parameter that controls the width of the peak. The full width at half maximum (FWHM) of a Gaussian peak can be calculated as $2\varphi_p\sqrt{2 \ln 2}$.

 Collisional broadening occurs due to collisions between particles, which effectively lower the characteristic time of the emission process. As a result of the uncertainty principle this increases the uncertainty in the energy of the emitted photons, which is described by a Lorentzian function:
\begin{equation} \label{eq:Lorentz}
f_L(\nu_j \mid \ell_p, \varphi_p) \propto \frac{\varphi_p^2}{(\nu_j - \ell_p)^2 + \varphi_p^2}
\end{equation}  
The FWHM of a Lorenztian peak is given by $2\varphi_p$. The heavier tails of the Lorentzian would imply long-range dependence between peaks. Failure to account for this would introduce bias, particularly if quantification was based on a single peak in isolation. Often the observed spectrum is the result of a combination of the above processes. This can be represented as a Voigt function, which is the convolution of a Gaussian and a Lorentzian \citep{Wertheim1974}.

%Semi-parametric, functional models such as these have previously been applied to other types of spectroscopy. \cite{Ritter1994} introduced a Bayesian model for the peaks in electron spectroscopy, using pseudo-Voigt functions for the broadening. They derived informative priors for the locations, FWHM, and mixing proportions of the peaks. This model was fitted to four peaks, removing the baseline as a pre-processing step. \cite{vanDyk2001,vanDyk2004} developed a joint model for the baseline and peaks in X-ray and $\gamma$-ray astronomy, using a Gaussian broadening function. This model is specifically suited to data with low photon counts, where the additive white noise assumption is inappropriate. 
% \citet{Kim2014} used a mixture model for detection of two-dimensional peaks in gas chromatography. 

%\cite{Fischer2002} and \cite{Razul2003} introduced reversible-jump Markov chain Monte Carlo (RJ-MCMC) algorithms \cite{Green1995} when the number of peaks is unknown. The baseline was estimated using a penalised spline, as in \citet{deRooi2012}, but the peaks were modelled as Gaussian functions \eqref{eq:GaussPk}. The number and locations of both the knots and the peaks were determined by the trans-dimensional algorithm. \citet{Wang2008} used RJ-MCMC to fit a similar model to mass spectrometry. These methods have only been applied to spectra where the baseline function is highly regular, or where the peaks are spaced far enough apart that the baseline can be directly observed. 

 Our main contribution is a Bayesian method for multivariate calibration (MVC). We extend the previous models of peaks and baselines in spectroscopy \citep[e.g.,][]{Ritter1994,vanDyk2001} to obtain estimates of the relationship between molecular concentration and amplitude for each peak. Our model  provides posterior distributions for quantities of scientific interest, such as the FWHM and LOD. We introduce a sequential Monte Carlo (SMC) algorithm \citep{DelMoral2006} to fit our model. We have implemented this algorithm as an open-source software package for the \textsf{R} statistical computing platform \citep{RCT2017}.

The remainder of this article is organised as follows. We present our Bayesian model and informative priors in Sect.~\ref{s:model}. Our SMC algorithm for fitting this model is described in Sect.~\ref{s:comp}. Results of applying our method to Raman spectroscopy are presented in Sect.~\ref{s:results}. We conclude the article with a discussion.

\section{Model}
\label{s:model}
We decompose a Raman spectrum into three major components:
\begin{equation}
\label{eq:model}
\bm{y}_i = \xi_i(\tilde{\bm\nu}) + c_i s(\tilde{\bm\nu}) + \bm\epsilon_i
\end{equation}
where $\bm{y}_i$ is a hyperspectral observation that has been discretised at a number of light frequencies or wavenumbers, $\nu_j \in \tilde{\bm\nu}$. The spectral signature $s(\tilde{\bm\nu})$ comprises the Raman peaks; $c_i$ is the concentration of the molecule; and $\xi_i(\tilde{\bm\nu})$ is the baseline. We assume that $\bm\epsilon_{\bullet}$ is zero mean, additive white noise with constant variance:
\begin{equation}\label{eq:noise}
\epsilon_{i,j} \sim \mathcal{N}\left( 0, \sigma^2_\epsilon \right)
\end{equation}
This assumption could be relaxed by allowing for autocorrelated residuals, as in \cite{Chib1993}. 

The baseline is a smoothly-varying, continuous function that is mainly due to background fluorescence. The main property that distinguishes the baseline from the other components of the signal is its smoothness. For this reason, we have chosen to model the baseline function as a penalised B-spline \citep{Eilers1996}:
\begin{equation}
\label{eq:spline}
\xi_i(\tilde{\bm\nu}) = \sum_{m=1}^M B_m(\tilde{\bm\nu}) \alpha_{i,m}
\end{equation}
where $B_m$ are the basis functions, $M$ is the total number of splines, and $\alpha_{i,\bullet}$ are the coefficients of the baseline for the $i$th observation. We use equally-spaced knots 10 cm$^{-1}$ apart, so that $M$ is typically $\approx$ 120. If the choice of knot locations is a concern, then a smoothing spline \citep{Eubank1999} could be used instead. %\citet{Razul2003} used an RJ-MCMC algorithm to determine the number and placement of the knots in the baseline function.

The Raman peaks are represented as an additive mixture of Gaussian, eq.~\eqref{eq:GaussPk}, or Lorentzian, eq.~\eqref{eq:Lorentz}, broadening functions:
\begin{equation} \label{eq:signature}
s(\nu_j) = \sum_{p=1}^P \beta_p \,
f_{\bullet}(\nu_j \mid \ell_p, \varphi_p),
\end{equation}
where $\beta_p$ is a regression coefficient that governs how the amplitude of the $p$th peak depends on the concentration of the molecule.
For the analysis in Sect~\ref{s:results}, we assume that the peak locations $\ell_p$ are known. However, a posterior distribution for the number of peaks $P$ and their locations could be obtained using the method in \cite{Harkonen2022}.

\begin{figure}
\centering
        \begin{subfigure}{0.84\textwidth}
\includegraphics[width=\textwidth]{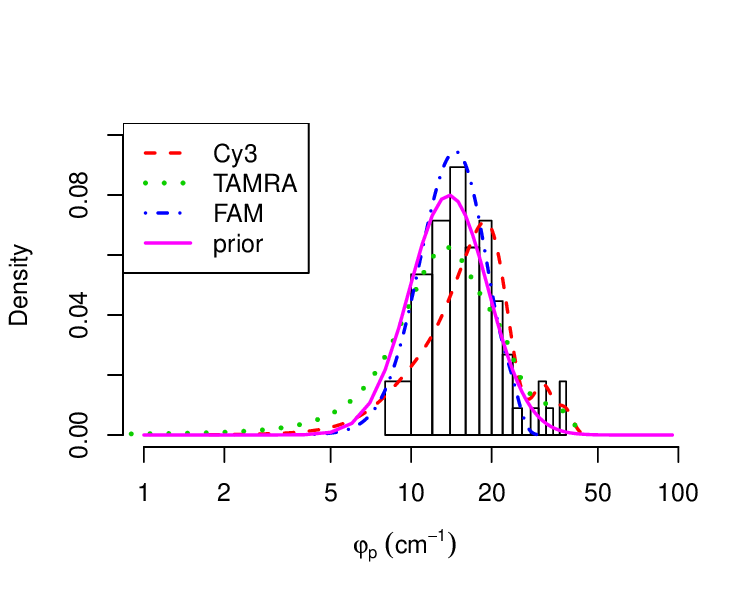}
\caption{Prior for Gaussian broadening.}\label{f:priorScaleGauss}
        \end{subfigure}%
\qquad
        \begin{subfigure}{0.84\textwidth}
\includegraphics[width=\textwidth]{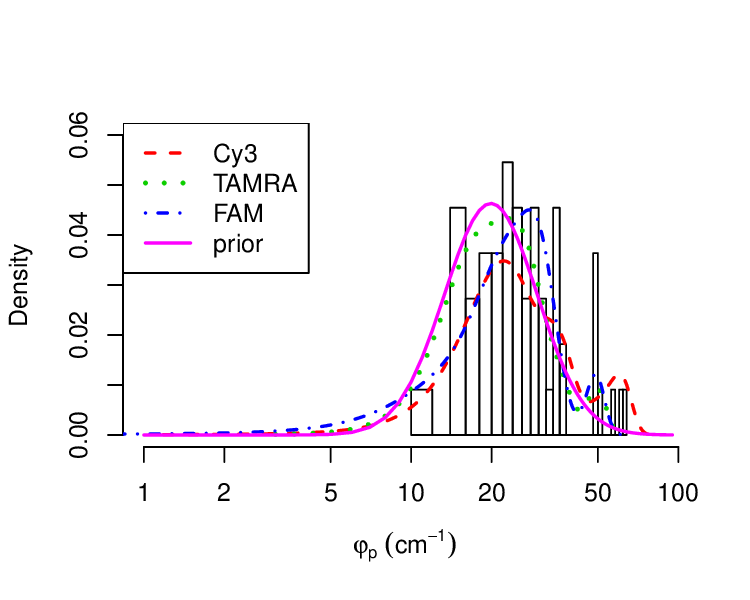}
\caption{Prior for Lorentzian broadening.}\label{f:priorScaleLorentz}
        \end{subfigure}%
\caption{Informative priors for the scale parameters of Raman peaks, derived from manual baseline correction and peak fitting of \textsf{Cy3}, \textsf{TAMRA} and \textsf{FAM} spectra using Grams/AI 7.00.}
\label{f:priors}
       \end{figure}
       
We derived informative priors for the scale parameters $\varphi_p$ by manual baseline correction and peak fitting in Grams/AI 7.00 (Thermo Scientific, Waltham, MA). We selected three representative spectra, one each of tetramethylrhodamine (\textsf{TAMRA}), fluorescein (\textsf{FAM}), and cyanine (\textsf{Cy3}), from an independent set of experimental data that had been previously analysed by \cite{Gracie2014}. We fitted both Gaussian and Lorentzian peaks to obtain the distributions shown in Fig.~\ref{f:priors}. A lognormal distribution  provided a good fit to the peaks in our training data. The median of the scale parameters was 16.47 for Gaussian peaks and the standard deviation of $\log\{\varphi_p\}$ was 0.34:
\begin{equation}
\label{eq:prior_scale}
\pi\left(\log\{\varphi_p\}\right) \sim  \mathcal{N}\left(\log(16.47) - \frac{0.34^2}{2}, \,0.34^2 \right)
\end{equation}
This agrees well with the theoretical value of 5 to 20 cm$^{-1}$ for broadening that is used in computational chemistry \citep{LeRu2009}. For Lorentzian peaks, the median was 25.27 and $\sigma\left(\log\{\varphi_p\}\right)$ was 0.4. These prior distributions overlap, although the Lorentzian peaks tend towards larger scale parameters. This is consistent with the FWHM, since rescaling the prior for the Gaussian peaks by $\sqrt{2 \log 2}$ results in a distribution that is very close to the prior for the Lorentzians.

\section{Computation}
\label{s:comp}
SMC algorithms evolve a population of weighted particles $\bm\Theta = \left[\bm\theta_q\right]_{q=1}^Q$ through a sequence of intermediate target distributions $\pi_0, \pi_1, \dots, \pi_T$. The particles are initialised from the joint prior distribution of the parameters:
\begin{equation}\label{eq:smcInit}
\pi_0(\bm\Theta) = \pi(\boldsymbol\beta) \pi(\bm\varphi) \pi\left(\bm\alpha \mid \sigma^2_\epsilon\right) \pi\left(\sigma^2_\epsilon\right) %\pi(\lambda \mid \sigma_\epsilon) \pi(\sigma_\epsilon)
\end{equation}
where each of these priors has been described in the previous section.

We use likelihood tempering SMC \citep{DelMoral2006} to fit our model to a single observed spectrum, $\bm{y}_i$. Under the assumption of additive Gaussian noise in eq.~\eqref{eq:model}, the likelihood is given by:
\begin{equation}\label{eq:like}
p(\bm{y}_i \mid \bm\Theta) = \prod_{\nu_j \in \tilde{\bm\nu}} \mathcal{N}\left( y_{i,j};\; \xi_i(\nu_j) + c_i s(\nu_j), \sigma^2_\epsilon \right)
\end{equation}
where the baseline function $\xi_i(\nu_j)$ is defined in eq.~\eqref{eq:spline} and the spectral signature $s(\nu_j)$ is defined in eq.~\eqref{eq:signature}. For multiple observations, 
we can update the posterior sequentially by combining this approach with iterated batch importance sampling \citep[IBIS;][]{Chopin2002}. These algorithms are implemented in our open-source \textsf{R} package `serrsBayes' \citep{serrsBayes}, which is available from the Comprehensive \textsf{R} Archive Network (CRAN) repository.

\section{Results}
\label{s:results}
In this section, we use our SMC algorithm to analyse a dilution study of \textsf{TAMRA} that was originally published in \cite{Gracie2014}. Metallic nanoparticles have been used to enhance the Raman signal, a technique known as surface-enhanced Raman scattering (SERS) \citep{LeRu2009}. Specifically, we used citrate-reduced silver nanoparticles (Ag NP) with mean diameter of 78 nm. SERS spectra have been obtained for 21 different concentrations, from 0.13 to 24.7 nanomolar (nM). There are 5 repeats of 3 technical replicates at each concentration, giving a total sample size of 315 spectra. These spectra were obtained using a 100 mW laser at 532 nm excitation wavelength. The resolution of the spectrometer was 0.5 cm$^{-1}$, providing 3601 measurements between 200 and 2000 cm$^{-1}$. Previously, \cite{Gracie2014} used univariate calibration to estimate a LOD of 99.5 picomolar (pM) for the peak at 1650 cm$^{-1}$. The aim of our analysis is to estimate the LOD of all 32 peaks simultaneously, using a Bayesian MVC approach.

\begin{table}
\caption{95\% highest posterior density (HPD) intervals for the regression coefficients $\beta_p$ (inverse nanomolar, nM$^{-1}$), full width at half maximum (FWHM, cm$^{-1}$) and limit of detection (LOD, nM) of the 18 largest peaks in the dilution study.}
\label{t:ibis_peaks}       % Give a unique label
\centering
\setlength{\tabcolsep}{12pt}
\begin{tabular}{rrrr}
$\ell_p$ (cm$^{-1}$) & $\beta_p$ (nM$^{-1}$) & FWHM (cm$^{-1}$)  & LOD (nM)\\
\noalign{\smallskip}\hline\noalign{\smallskip}
460 & [6.73; 17.23] & [0.00; 14.43] & [0.211; 0.784] \\
   505 & [167.95; 177.83] & [14.36; 15.52] & [0.019; 0.047] \\
   632 & [253.65; 263.16] &  [11.54; 12.13] &  [0.012; 0.032]\\
   725 & [19.63; 29.52]  & [9.97; 16.43] &  [0.123; 0.316]\\
   752 & [44.74; 54.81]  & [19.49; 23.45] &  [0.059; 0.156]\\
   843 & [23.48; 33.73]  & [15.97; 22.26] &  [0.106; 0.297]\\
   965 & [17.78; 28.09]  & [12.07; 20.32] &  [0.135; 0.355]\\
   1140 & [25.49; 36.11]  & [20.41; 27.92] & [0.089; 0.253]\\
   1190 & [18.67; 28.99]  & [11.27; 19.36] & [0.110; 0.352]\\
   1220 & [147.84; 158.30]  & [17.68; 19.20] & [0.020; 0.051]\\
   1290 & [31.19; 42.49]  & [15.72; 25.80]  & [0.080; 0.213]\\
   1358 & [210.46; 221.96]  & [17.63; 18.97] & [0.015; 0.035]\\
   1422 & [53.13; 63.99]  & [15.34; 18.16] & [0.059; 0.135]\\
   1455 & [39.05; 53.87]  & [20.61; 41.11] & [0.069; 0.175]\\
   1512 & [146.07; 157.28]  & [20.81; 22.38] & [0.019; 0.049]\\
   1536 & [209.18; 221.03]  & [14.43; 15.80] & [0.014; 0.036]\\
   1570 & [38.54; 53.67]  & [23.87; 51.02]  & [0.073; 0.173]\\
   1655 & [467.58; 477.91] & [17.40; 17.92] & [0.006; 0.016]\\
\noalign{\smallskip}\hline\noalign{\smallskip}
\end{tabular}
\end{table}

\begin{figure}
\centering
        \begin{subfigure}{\textwidth}
\includegraphics[width=\textwidth]{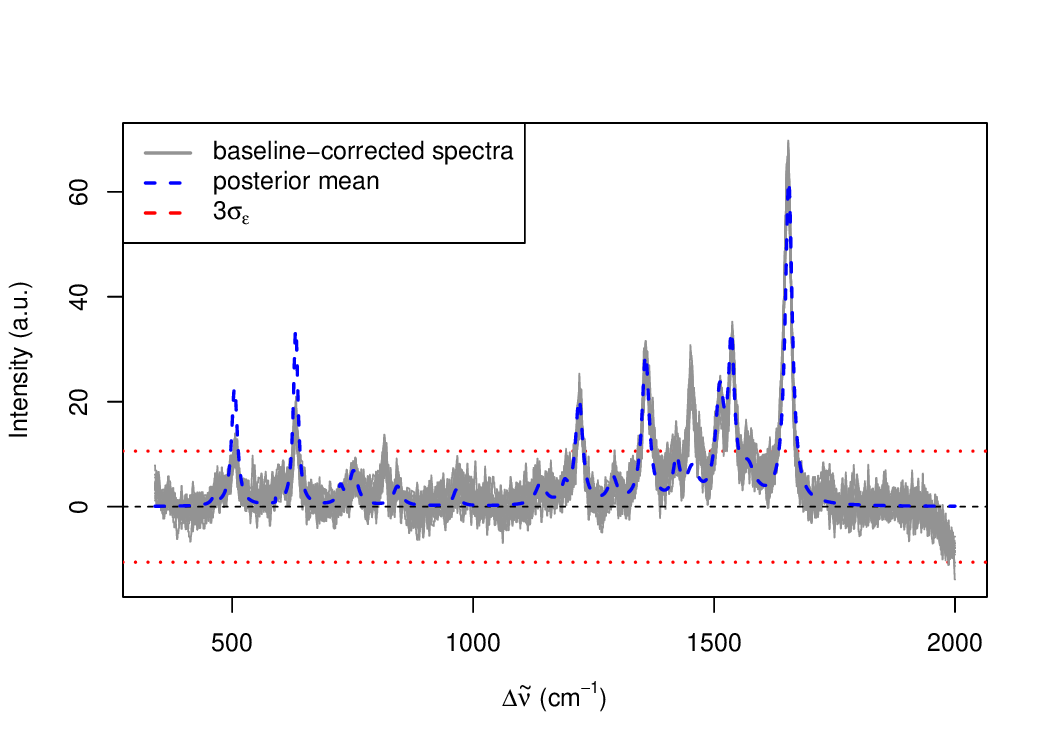}
\caption{\textsf{TAMRA} at 0.13 nM.} \label{f:TAMRAconc013}
        \end{subfigure}%
\qquad
        \begin{subfigure}{\textwidth}
\includegraphics[width=\textwidth]{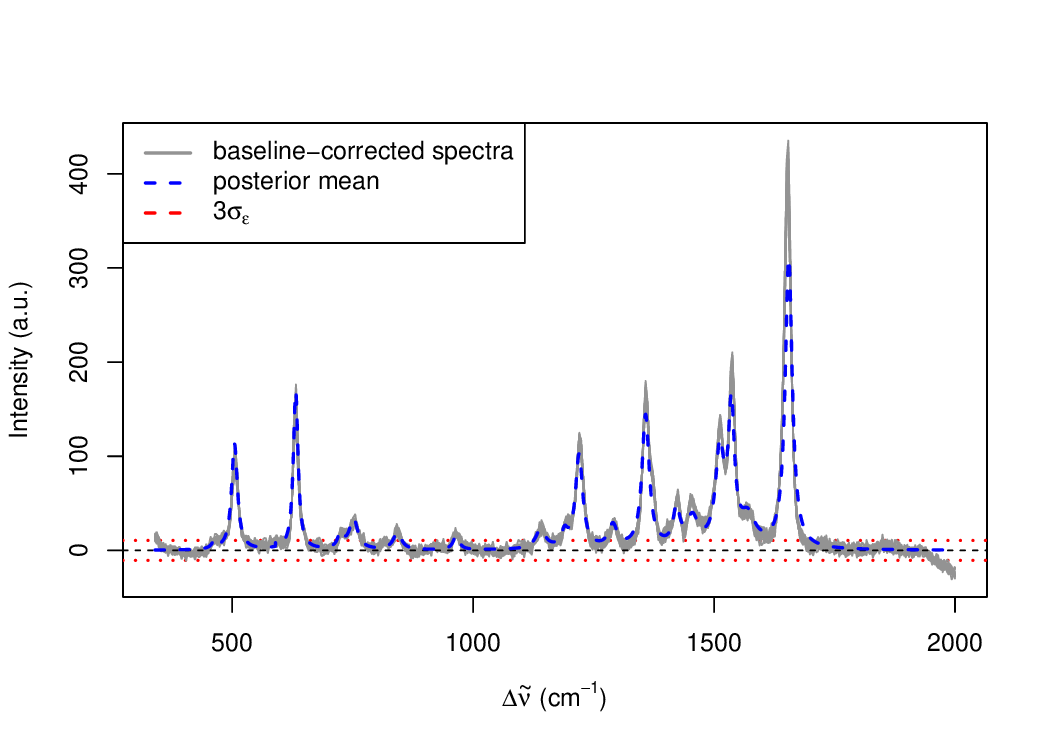}
\caption{\textsf{TAMRA} at 0.65 nM.} \label{f:TAMRAconc065}
        \end{subfigure}%
\caption{Surface-enhanced Raman scattering (SERS) spectra and model fit at very low  concentrations of tetramethylrhodamine (\textsf{TAMRA}) dye.} \label{f:lowConc}
\end{figure}

95\% highest posterior density (HPD) intervals for the regression coefficients $\beta_p$, the FWHM and the LOD of the 18 largest peaks are shown in Table~\ref{t:ibis_peaks}.
To verify the HPD intervals for the LOD, we can closely examine the spectra at the two lowest concentrations, 0.13 and 0.65 nM. The lower bounds for detectability of the peaks at 460 and 965 cm$^{-1}$ are greater than 0.13 nM, so we would not expect those peaks to be visible at that concentration. Conversely, the upper bounds for 7 of the peaks are lower than 0.13 nM, so we would expect all of those peaks to be clearly visible, as shown by Fig.~\ref{f:TAMRAconc013}. There is also an eighth peak at 1455 cm$^{-1}$ that has been underestimated by the model. The upper bounds for all of the peaks except at 460 cm$^{-1}$ are lower than 0.65 nM, so at least 17 out of 34 peaks should be visible at that concentration, as shown in Fig~\ref{f:TAMRAconc065}. Care must be taken when extrapolating beyond the range of the data, but we predict that overall the LOD for \textsf{TAMRA} is between 6 and 16 picomolar (pM). 

\section{Conclusion}
\label{s:conc}
This study has presented a robust Bayesian framework for multivariate calibration (MVC) in spectroscopy. Our model-based approach provides several advantages over existing quantitative methods, which rely on baseline subtraction as a separate preprocessing step. 
Employing a sequential Monte Carlo (SMC) algorithm, we have demonstrated that it is possible to jointly estimate the parameters of the spectral signature and the smoothly-varying baseline. Our integrated Bayesian approach provides well-calibrated quantification of uncertainty, including 95\% highest posterior density (HPD) intervals. 
We can directly estimate quantities of scientific interest, such as the amplitudes, limit of detection (LOD), and full width at half maximum (FWHM) of the Raman peaks. We have implemented our algorithm as an open-source \textsf{R} package \citep{serrsBayes}, which represents an important tool for analysing experimental data.

Our model could be extended to perform detection and quantification of multiplexed spectra, where several dye molecules may be present. Such a model would need to account for nonlinear interactions between molecules, for example due to preferential attachment \citep{Gracie2015}. Estimating the LOD and the limit of quantification (LOQ) for each peak would be particularly useful in this setting, since many of the peaks of different molecules overlap with each other. Such estimates could be used in experimental design, to select molecules that maximise differentiation between their spectral signatures. It would also be useful to extend our model to include spatial correlation between spectra. Some spectrometers are able to collect measurements on a 2D or 3D lattice, known as a Raman map. A divide-and-conquer SMC approach \citep{Lindsten2017} or Bayesian neural network \citep{harkonen2024} for high-throughput computation could be applied in this setting, to process the large volumes of data that are involved.

\section*{Acknowledgements}
This work was funded by the UK EPSRC programme grant, ``{\em In Situ Nanoparticle
Assemblies for Healthcare Diagnostics and Therapy}'' (ref: EP/L014165/1) and an
Award for Postdoctoral Collaboration from the EPSRC Network on Computational
Statistics \& Machine Learning (ref: EP/K009788/2).
%\end{acknowledgement}
%

\bigskip
%
% BibTeX users please use
\bibliographystyle{plainnat}
\bibliography{serrs}
\end{document}